# Reconfigurable photonic RF filters based on integrated Kerr frequency comb sources

Xingyuan Xu,[1] Mengxi Tan,[1] Jiayang Wu,[1] Thach G. Nguyen,[2] Sai T. Chu,[3] Brent E. Little,[4] Roberto Morandotti,[5,6,7] Arnan Mitchell,[2] and David J. Moss[1, *]

[1]Centre for Micro-Photonics, Swinburne University of Technology, Hawthorn, VIC 3122, Australia
[2]School of Engineering, RMIT University, Melbourne, VIC 3001, Australia
[3]Department of Physics and Material Science, City University of Hong Kong, Tat Chee Avenue, Hong Kong, China.
[4]Xi'an Institute of Optics and Precision Mechanics Precision Mechanics of CAS, Xi'an, China.
[5]INRS-Énergie, Matériaux et Télécommunications, 1650 Boulevard Lionel-Boulet, Varennes, Québec, J3X 1S2, Canada.
[6]ITMO University, St. Petersburg, Russia.
[7]Institute of Fundamental and Frontier Sciences, University of Electronic Science and Technology of China, Chengdu 610054, China.
*dmoss@swin.edu.au

*Abstract*—We demonstrate two categories of photonic radio frequency (RF) filters based on integrated optical micro-combs. The first one is based on the transversal filtering structure and the second one is based on the channelization technique. The large number of wavelengths brought about by the microcomb results in a significantly increased RF spectral resolution and a large instantaneous bandwidth for the RF filters. For the RF transversal filter, we demonstrated Q factor enhancement, improved out-of-band rejection, tunable centre frequency, and reconfigurable filtering shapes. While a high resolution of 117 MHz, a large RF instantaneous bandwidth of 4.64 GHz, and programmable RF transfer functions including binary-coded notch filters and RF equalizing filters with reconfigurable slopes are demonstrated for the RF channelized filter. The microcomb-based approaches feature a potentially much smaller cost and footprint, and is promising for integrated photonic RF filters.

## I. Introduction

The performance of modern radar and communications systems is mainly determined by their capabilities in radio frequency (RF) signal processing [1]. Photonic RF signal processors have attracted great interest since they can provide wide RF bandwidths [2-4], in contrast to their electrical counterpart that are subject to the so called "electronic bandwidth bottleneck". As a basic function of signal processing, RF filters with broad operational bandwidths, high reconfigurability, high frequency selectivity, and low cost are highly desired.

Extensive efforts have been made to achieve photonic RF filters. Generally, photonic RF filters can be separated into two categories. The first includes optical filters that map their optical transmission response onto the RF domain, readily achieving narrow bandwidths down to the MHz level [5], although facing challenges in achieving reconfigurable transfer functions. The other category featuring transversal structures [6-12] can achieve arbitrary RF transfer functions via changing the tap weights, where Bragg grating arrays [9], discrete laser arrays [10], or electro-optic combs have been employed to establish the needed taps [11, 12]. However, although offer advantages, these approaches were still subject to limitations in one form or another, such as the significantly increased complexity and reduced performance due to the limited number of available taps, or the need of high-frequency RF sources.

Integrated Kerr micro-comb sources [13-22], originating from the parametric oscillation in on-chip micro-ring resonators [23, 24], offer distinct advantages over traditional multi-wavelength sources for transversal RF filters, such as the potential to provide a much higher number of wavelengths as well as greatly reduced footprint and complexity. Advanced RF functions based on Kerr micro-combs [25-32] have achieved a high degree of versatility and dynamic reconfigurability.

In this paper, we first report a microcomb-based photonic RF transversal filter with 80 taps, enabled by a 49GHz-free-spectral-range integrated Kerr micro-comb source. This resulted in a Q factor for the RF bandpass filter of four times higher than previous results [32]. Further, by programming the spectral shape of the Kerr optical micro-comb, we achieve RF filters with a high out-of-band rejection of up to 48.9 dB using Gaussian apodization, as well as a significantly improved tunable centre frequency covering over 90% of the RF Nyquist band. Further, by using the on-chip micro-comb in combination with a passive MRR with a free spectral range (FSR) of ~49 GHz and a Q factor of over 1.5 million, we realize a high-resolution photonic RF filter with a large number of wavelength channels (80 over the C-band and > 160 over the C/L-bands) that enabled a large operation bandwidth. We experimentally demonstrated programmable arbitrary transfer functions for RF spectral shaping with an operation bandwidth of 4.64 GHz, together with a high resolution of 117 MHz. Our experimental results agree well with theory, verifying the feasibility of our approach towards the realization of high performance advanced RF filters with potentially reduced cost, footprint, and complexity than other solutions.

## II. Microcomb generation

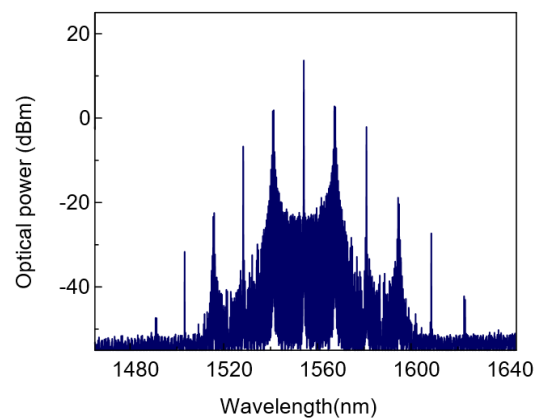

Fig. 1. Optical spectrum of the generated soliton crystal micro-comb.

The MRR used to generate the Kerr optical micro-comb was fabricated on a high-index doped silica glass platform using CMOS-compatible fabrication processes. First, high-index (n = ~1.7 at 1550 nm) doped silica glass films were deposited using plasma enhanced chemical vapour deposition, then patterned by deep ultraviolet photolithography and etched via reactive ion etching to form waveguides with exceptionally low surface roughness. Finally, silica (n = ~1.44 at 1550 nm) was deposited as an upper cladding. The advantages of our platform for optical micro-comb generation include ultra-low linear loss (~0.06 dB cm$^{-1}$), a moderate nonlinear parameter (~233 W$^{-1}$ km$^{-1}$), and in particular a negligible nonlinear loss up to extremely high intensities (~25 GW cm$^{-2}$). Due to the ultra-low loss of our platform, the MRR features narrow resonance linewidths that corresponds to a quality factor of ~1.5 million. The radius of the MRR was ~592 µm, corresponding to an optical free spectral range of ~0.4 nm or ~49 GHz. The small optical free spectral range of the MRR enabled 80 wavelengths, or taps for the transversal filter, in the C band.

To generate Kerr micro-combs, we adjusted the polarization and wavelength of the pump light to one of the TE resonances of the MRR at ~1553.2 nm with the pump power set at ~30.5 dBm. When the detuning between the pump wavelength and the cold resonance became small enough, such that the intra-cavity power reached a threshold value, modulation instability driven oscillation was initiated. As the detuning was changed further, distinctive 'fingerprint' optical spectra were observed (Fig. 1), which arose from the interference of co-circulating solitons in the MRR that were termed as "soliton crystals".

### III. 80-TAP TRANSVERSAL RF FILTER

The transfer function of a photonic transversal filter can be described as

$$H(\omega) = \sum_{n=0}^{N-1} h(n) e^{-j\omega nT} \quad (1)$$

where $\omega$ is the angular frequency of the input RF signal, $N$ is the number of taps, $h(n)$ is the discrete impulse response representing the tap coefficient of the $n_{th}$ tap, and $T$ is the time delay between adjacent taps. The free spectral range of the transversal filter $FSR_{RF}$ is given by $1/T$. By properly setting the tap coefficients ($h(n)$, $n = 0, 1, …, N-1$) for different spectral transfer functions, a reconfigurable transversal filter with arbitrary filter shapes can be achieved.

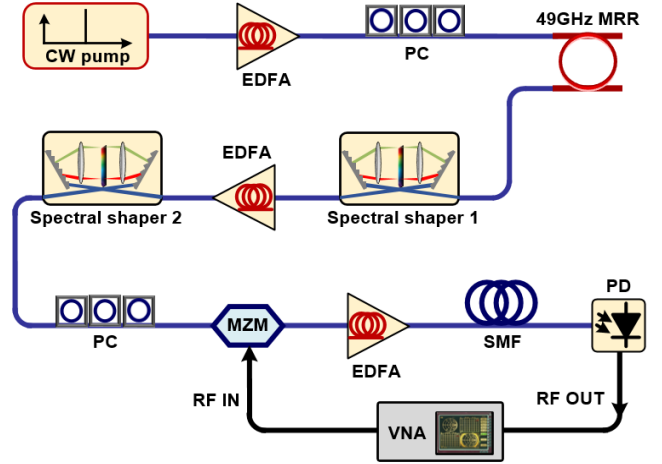

Fig. 2. Schematic diagram of the 80-tap photonic RF transversal filter based on an integrated 49GHz-spacing micro-comb source. EDFA: erbium-doped fiber amplifier. PC: polarization controller. MRR: micro-ring resonator. WS: Waveshaper. MZM: Mach-Zehnder modulator. SMF: single mode fiber. OC: optical coupler. OSA: optical spectrum analyzer. PC: computer. PD: photodetector. VNA: vector network analyzer.

Figure 2 shows a schematic diagram of the 80-tap photonic RF transversal filter based on an integrated Kerr micro-comb source. The generated Kerr micro-comb served as a multi-wavelength source where the power of each comb line was manipulated by the Waveshapers to achieve the designed tap weights. The shaped comb lines were then fed into an EO intensity modulator, yielding replicas of the input RF signal in the optical domain. The modulated signal produced by the intensity modulator went through a spool of dispersive fiber, generating a time delay *T* between adjacent taps. Finally, the weighted and delayed taps were combined upon photo detection and converted back into RF signals at the output.

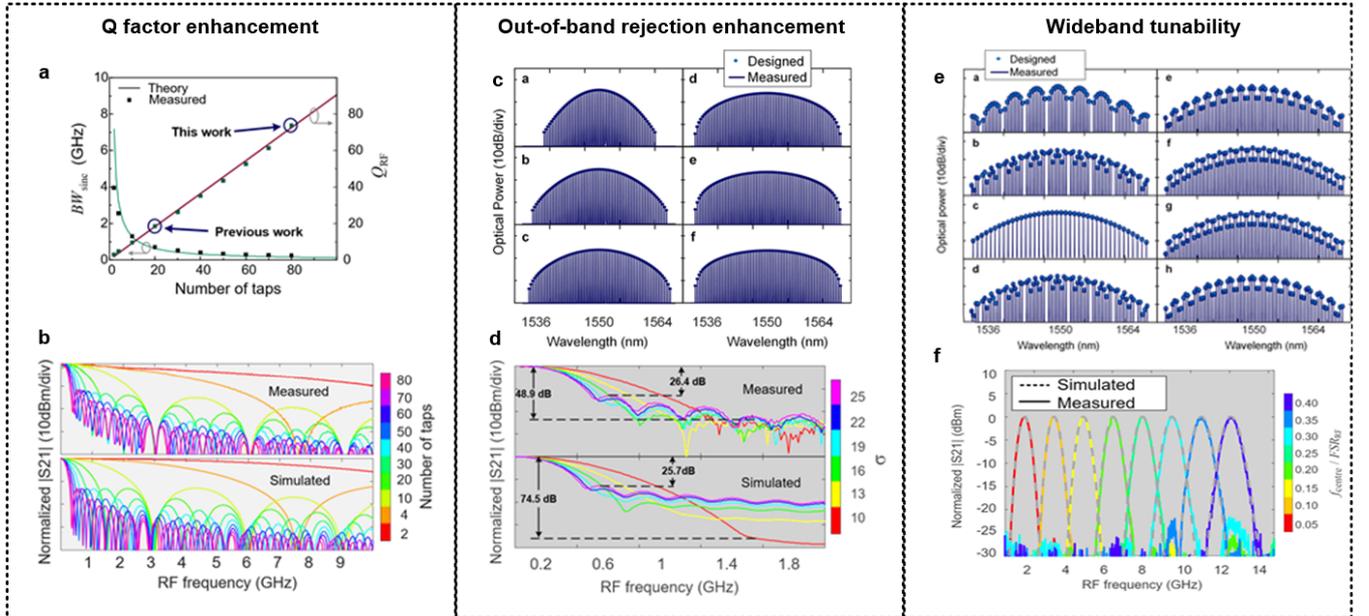

Fig. 3. (a) Calculated relationship between $Q_{RF}$ factor and the number of taps. (b) RF transmission spectra of the sinc filter with different tap numbers. (c) Optical spectra of the shaped micro-comb corresponding to the Gaussian-apodized sinc filter. (d) RF transmission spectra of Gaussian apodized sinc filter. (e) Optical spectra of the shaped micro-comb corresponding to the centre-frequency-tunable filter. (f) RF transmission spectra of the centre-frequency-tunable filter.

To demonstrate the $Q_{RF}$ factor enhancement brought about by the microcomb's large number of wavelengths, we implemented a low-pass sinc filter (i.e., $h_{sinc}(n)=1$) with different numbers of taps. The 3-dB bandwidth ($BW_{sinc}$) was measured to calculate the corresponding $Q_{RF}$ factor ($Q_{RF} = BW_{sinc} / FSR_{RF}$). The RF transmission spectra of the sinc filter (Fig. 3(b)), measured by a vector network analyser (VNA, Anritsu 37369A), showed good agreement with theory (Fig. 3(a)). The measured $BW_{sinc}$ decreased from 3.962 to 0.236 GHz when the tap number was increased from 2 to 80, indicating a greatly enhanced $Q_{RF}$ of up to 73.7 with 80 taps.

To improve the out-of-band rejection of the transversal filter, a Gaussian apodization was applied to the sinc filter, with a main-to-secondary sidelobe ratio of up to 48.9 dB (Fig. 3(c, d)). In order to demonstrate the tunability of the 80-tap transversal filter's centre frequency, the tap coefficients of the Gaussian-apodized sinc filter were multiplied by a sine function to shift the RF transmission spectrum. The corresponding RF transmission spectra (Fig. 3(f)) shows a tunable centre frequency ranging from $0.05 \times FSR_{RF} = 1.4$ GHz to $0.40 \times FSR_{RF} = 11.5$ GHz with a relatively high MSSR of >25 dB, occupying 90% of the Nyquist band.

In addition, we also demonstrated channelized RF filters based on a passive MRR (Fig. 4). The input RF signal was multi-cast onto each comb line via a phase modulator and fed to the passive MRR for spectral slicing. The TM mode through-port transmission of the passive MRR was employed to perform phase-to-intensity modulation conversion by filtering out the lower sideband and, at the same time, map its high-Q resonances (notches) onto the RF domain for high-resolution RF filtering. To demonstrate arbitrary RF transfer functions, we varied the channel weights and measured the RF transmission spectra with a vector network analyser. Figure 5 shows a reconfigurable notch filter achieved with binary (i.e., either "1" or "0") channel weights. The RF transmission spectra match very well with the shaped comb spectra, proving the reconfigurability and resolution of our photonic RF filter.

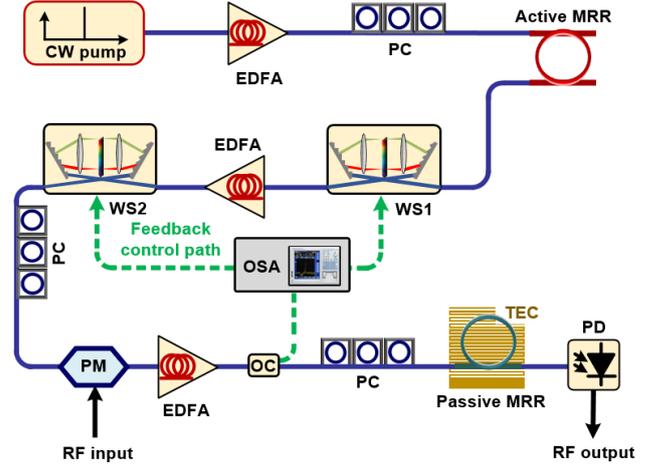

Fig. 4. Schematic diagram of the microcomb-based RF filter. EDFA: erbium-doped fiber amplifier. PC: polarization controller. MRR: micro-ring resonator. WS: Waveshaper. PM: Phase modulator. OC: optical coupler. OSA: optical spectrum analyzer. TEC: Thermoelectric cooler. PD: photodetector.

## IV. Conclusions

We demonstrate microcomb-based photonic RF transversal filters with 80 taps and channelized RF filters. The large number of wavelengths brought about by the microcomb results in a significantly increased RF spectral resolution and a large instantaneous bandwidth for the RF filters. For the RF transversal filter, we demonstrated Q factor enhancement, improved out-of-band rejection, tunable centre frequency, and reconfigurable filtering shapes. While a high resolution of 117 MHz, a large RF instantaneous bandwidth of 4.64 GHz, and programmable RF transfer functions including binary-coded notch filters and RF equalizing filters with reconfigurable slopes are demonstrated for the RF channelized filter. The microcomb-based approaches feature a potentially much smaller cost and footprint, and is promising for integrated photonic RF filters.

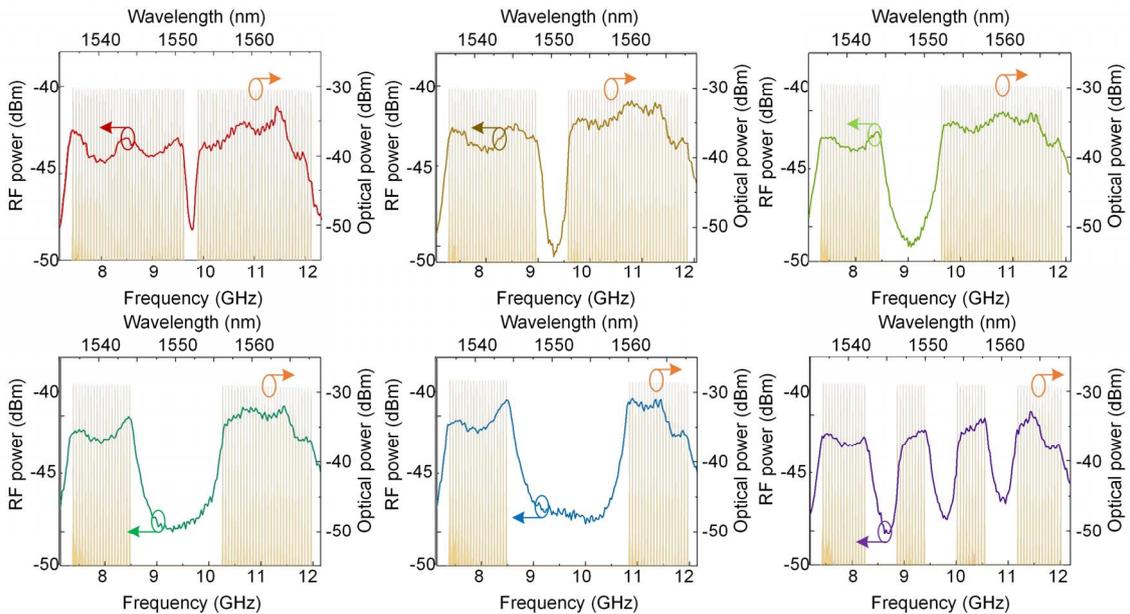

Fig. 5. Measured optical spectra of the shaped micro-comb and corresponding RF transmission spectra of the RF spectral shaper featuring notches with varying bandwidth.

ACKNOWLEDGMENT

This work was supported by the Australian Research Council Discovery Projects Program (No. DP150104327). RM acknowledges support by the Natural Sciences and Engineering Research Council of Canada (NSERC) through the Strategic, Discovery and Acceleration Grants Schemes, by the MESI PSR-SIIRI Initiative in Quebec, and by the Canada Research Chair Program. He also acknowledges additional support by the Government of the Russian Federation through the ITMO Fellowship and Professorship Program (grant 074-U 01) and by the 1000 Talents Sichuan Program in China. Brent E. Little was supported by the Strategic Priority Research Program of the Chinese Academy of Sciences, Grant No. XDB24030000.